\documentclass[aps,prc,groupedaddress,twocolumn,showpacs,amsmath,amssymb]{revtex4}
\usepackage{graphicx}
\usepackage{dcolumn}
\usepackage{hyperref}
\usepackage{bm}

\begin{document}

\title{Effect of $\Delta$ potential on the $\pi^-/\pi^+$ ratio in heavy-ion collisions at intermediate energies}

\author{Wen-Mei Guo$^{1,2,3}$}
\author{Gao-Chan Yong$^{1,4}$}\email{yonggaochan@impcas.ac.cn}
\author{Wei Zuo$^{1,4}$}

\affiliation{%
$^1${Institute of Modern Physics, Chinese Academy of Sciences, Lanzhou 730000, China}\\
$^2${School of Physical Science and Technology, Lanzhou University,
Lanzhou 730000, China}\\
$^3${University of Chinese Academy of Sciences, Beijing 100049, China}\\
$^4${State Key Laboratory of Theoretical Physics, Institute of
Theoretical Physics, Chinese Academy of Sciences, Beijing, 100190}\\
}%

\date{\today}

\begin{abstract}
Based on the isospin-dependent Boltzmann-Uehling-Uhlenbeck(IBUU) transport model,
effects of $\Delta$ resonance potential on the free n/p and $\pi^-/\pi^+$ ratios in the central collision
of $^{197}Au+^{197}Au$ at beam energies of 200 and 400 MeV/A are studied.
It is found that the effect of $\Delta$ potential on the ratio of pre-equilibrium free $n/p$ is invisible. The effect of $\Delta$ isovector potential on the kinetic energy integrating ratio of $\pi^-/\pi^+$ may be observable only at lower incident beam energies and with stiffer symmetry energy.
The strength of the $\Delta$ isoscalar potential affects the height of the $\pi^-/\pi^+$ ratio around the Coulomb peak but does not affect the kinetic energy integrating ratio of $\pi^-/\pi^+$. In heavy-ion collisions at intermediate energies, relating to the question of non-conservation of energy on
$\Delta$ or $\pi$ productions, one can replace the $\Delta$ potential by nucleon isoscalar potential especially a soft symmetry energy is employed.

\end{abstract}


\maketitle

\section{MOTIVATIONS}

In recent years, the research of the density dependent symmetry energy is still one of the hot topics in nuclear physics and astrophysics communities. This is simply because the symmetry energy governs many nuclear and astrophysical phenomena, such as the cooling of neutron stars \cite{JMLatt91}, the mass-radius relations of neutron stars \cite{MPrak88}, the study of Gravitational waves \cite{FJFatt14}, properties of nuclei involved in r-process nucleosynthesis \cite{NNik11}, and observables in heavy-ion collisions at intermediate energies \cite{Liba97,MBTSA09}. While significant progress has been made in constraining the symmetry energy around saturation density \cite{MBTSA09,MBTsa11,YXZhang13,KSJeo13}, it is rather uncertain
at supra-saturation densities. Fortunately, more related experiments are currently underway or being planed at several advanced radioactive ion beam facilities such as CSR/China \cite{ZGX09}, FRIB/USA \cite{USA}, GSI/Germany \cite{GSI}, RIKEN/Japan \cite{RIKEN}, or KoRIA/Korea \cite{Korea}. It thus provides more opportunities and enhances our confidences to study the symmetry energy although there are still many uncertain factors, such as the effects of pion potential \cite{WMGuo15,hongj2014}, the isospin dependence of in-medium nuclear strong interactions \cite{yong2011}, the short-range tensor force \cite{VRP72,xuc2011} and the $\pi-N-\Delta$ dynamics \cite{ko2014,ditoro2005,cozma2014}. 

It is noted that 
the sensitivities of the $\pi^-/\pi^+$ ratio to the symmetry energy shown in Ref.~\cite{hongj2014,ko2014} is quite different from that shown in Ref.~\cite{WMGuo15}. The discrepancies observed among the results of different models may be because the latter uses a momentum-dependent nucleon symmetry potential while the previous two studies use a momentum-independent nucleon symmetry potential. The momentum-dependent nucleon symmetry potential could increase the sensitivity of dense neutron to proton ratio (thus the $\pi^-/\pi^+$ ratio) to the symmetry energy \cite{bcsc2003}. While in this study, we use a modified symmetry potential which including the effects of the short-range correlations of neutron and proton \cite{OHen15}. The short-range correlations of neutron and proton reduce the kinetic symmetry energy, the strength of the modified symmetry potential is thus increased \cite{OHen15}. The effects of such modified symmetry potential on the $\pi^-/\pi^+$ ratio may be also enlarged. In the following studies, we confirmed the above deduction.

In Refs.~\cite{ko2014,ditoro2005,cozma2014} it is argued that due to the difference of $\Delta$ potential and nucleon potential, threshold conditions of pion and $\Delta$ productions need to be modified. Relating to the question of non-conservation of energy on
$\Delta$ or $\pi$ productions, in this study, we also investigate how the difference of $\Delta$ and nucleon potentials affects pion production. If the effect of $\Delta$ potential on $\pi^-/\pi^+$ ratio can be neglected, one can use nucleon potential to replace $\Delta$ potential. Our study shows that one can replace the $\Delta$ potential by the nucleon isoscalar potential if a soft symmetry energy is employed.

\section{The IBUU Model and the potential of $\Delta$ resonance}

\begin{figure*}[htb]
\centering\emph{}
\includegraphics[width=0.99\textwidth]{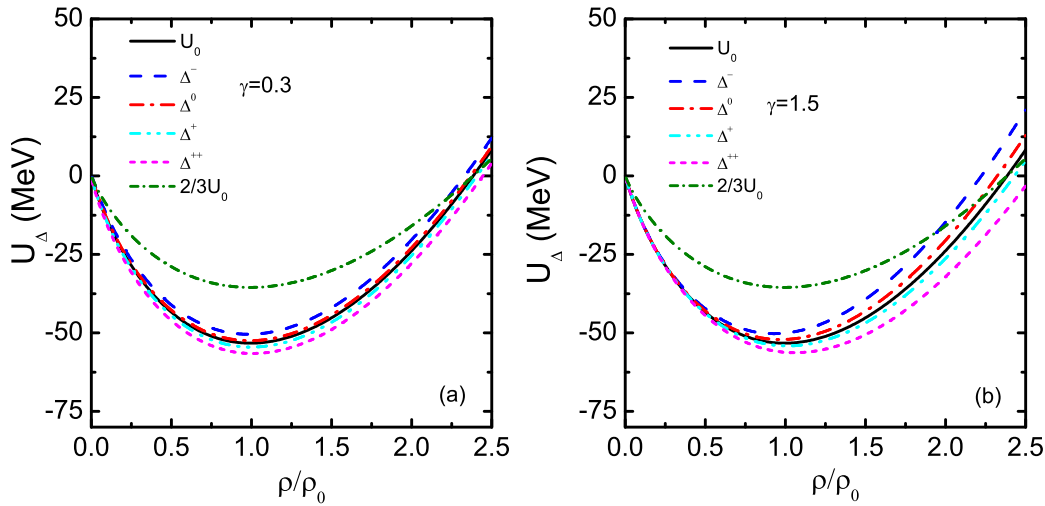}
\caption{(Color online) Density-dependent $\Delta$ resonance potential. The left window shows  $\Delta$ resonance potential with the soft ($\gamma=0.3$) symmetry potential while right window shows that with the stiff ($\gamma=1.5$) symmetry potential ($\delta$ = 0.2).} \label{Deltapot}
\end{figure*}
In this study, we use the semi-classical transport model IBUU, in which a momentum-independent and isospin-dependent mean-field  potential is adopted \cite{OHen15}. The symmetry potential reads
\begin{eqnarray}
U_{sym}(n/p)&=&[E_{sym}(\rho_0)-E^{kin}_{sym}(\rho_0)](\frac{\rho}{\rho_0})^{\gamma}\nonumber\\
& &\times[\pm2\frac{\rho_n-\rho_p}{\rho}+(\gamma-1)(\frac{\rho_n-\rho_p}{\rho})^2],
\end{eqnarray}
where $E_{sym}(\rho_0)$ = 31.6 MeV is the symmetry energy at saturation density, $E^{kin}_{sym}(\rho_0)$ = -6.71 MeV is the kinetic symmetry energy at saturation density, and $\gamma$ denotes the degree of stiffness of the symmetry potential.

The isoscalar potential for $\Delta$ resonance has been studied by many people using various many-body approaches and interactions \cite{EOset81,MBal90,GEB75,TEric88}. Considering that the mean field potential of $\Delta$ resonance mainly depends on its interactions with surrounding nucleons, one can make an assumption that the isoscalar potential of $\Delta$ resonance is the same as that for nucleon. The isovector potential for $\Delta$ resonance can be an weighted average of that for
neutrons and protons according to the charge state of $\Delta$ resonance in the processes $\Delta \leftrightarrow \pi N$ \cite{ko2014,ditoro2005,cozma2014,BALi12}. It reads
\begin{eqnarray}
U({\Delta^-})&=& U(n),\\
U({\Delta^0})&=& \frac{2}{3}U(n)+\frac{1}{3}U(p),\\
U({\Delta^+})&=& \frac{1}{3}U(n)+\frac{2}{3}U(p),\\
U({\Delta^{++}})&=& U(p).
\end{eqnarray}
For comparison, we also adopt the potential of $\Delta$ resonance to be equal to the isoscalar potential of nucleon as the other choice, i.e.,
\begin{eqnarray}
U_{\Delta}=U_0.
\end{eqnarray}
By phenomenology $\Delta$ potential has a depth of about -30 MeV at $\rho_0$ while nucleon potential is approximately -50 MeV deep \cite{TEric88}. Therefore, one can also use the form \cite{OBuss08}
\begin{eqnarray}
U_{\Delta}=\frac{2}{3}U_0.
\end{eqnarray}
Shown in Fig.~\ref{Deltapot} is our used density-dependent potentials of $\Delta$ resonance with soft ($\gamma=0.3$) and stiff ($\gamma=1.5$) symmetry potentials, respectively. One can see that the density-dependent $\Delta$ potential may also depend on the symmetry potential used.

\section{Results and discussions}

\begin{figure}[t!]
\begin{center}
\includegraphics[width=0.5\textwidth]{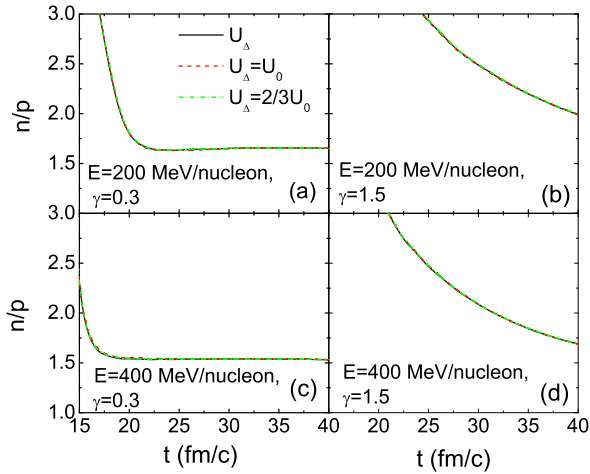}
\end{center}
\caption{(Color online) Time evolution of pre-equilibrium free $n/p$ ratio with different forms of $\Delta$ potential and nucleon symmetry potential ($\gamma=0.3,\  \gamma=1.5$) in  $^{197}Au+^{197}Au$ reaction at beam energies of 200 and 400 MeV/A, respectively.} \label{npt}
\end{figure}
\begin{figure}[t!]
\begin{center}
\includegraphics[width=0.5\textwidth]{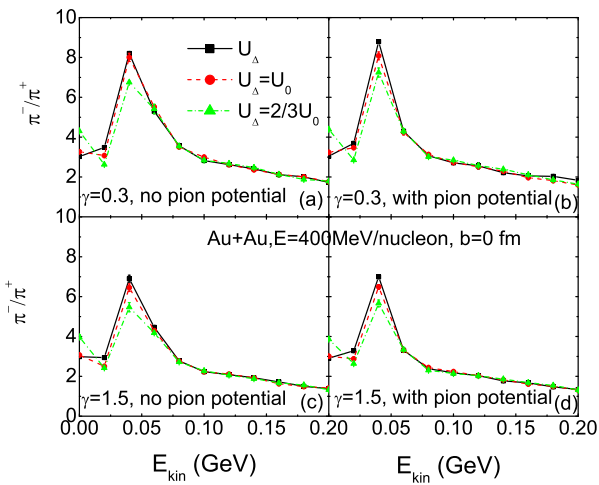}
\end{center}
\caption{(Color online) Kinetic energy dependence of $\pi^-/\pi^+$ ratio with different forms of $\Delta$ and nucleon symmetry potential ($\gamma=0.3,\  \gamma=1.5$) with or without pion potentials in $^{197}Au+^{197}Au$ reaction at a beam energy of 400 MeV/A.} \label{E400piEk}
\end{figure}
\begin{figure}[t!]
\begin{center}
\includegraphics[width=0.5\textwidth]{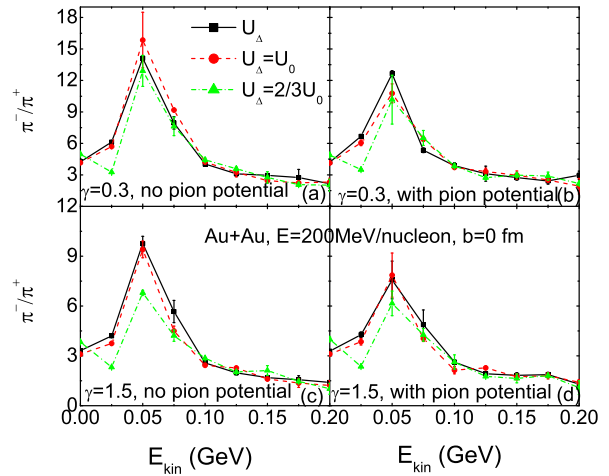}
\end{center}
\caption{(Color online) Same as Figure~\ref{E400piEk}, but for the beam energy of 200 MeV/A.} \label{E200piEk}
\end{figure}
\begin{figure}[htb]
\begin{center}
\includegraphics[width=0.5\textwidth]{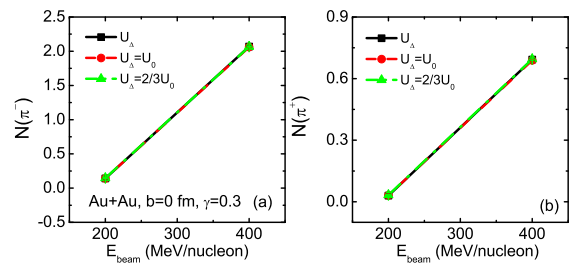}
\end{center}
\caption{(Color online) Effects of different forms of $\Delta$ resonance potential on $\pi^-$ and $\pi^+$ productions with soft symmetry potential ($\gamma=0.3$) in central $^{197}Au+^{197}Au$ collision at beam energies of 200 and 400MeV/A, respectively.} \label{g03-Npi}
\end{figure}
\begin{figure}[htb]
\begin{center}
\includegraphics[width=0.5\textwidth]{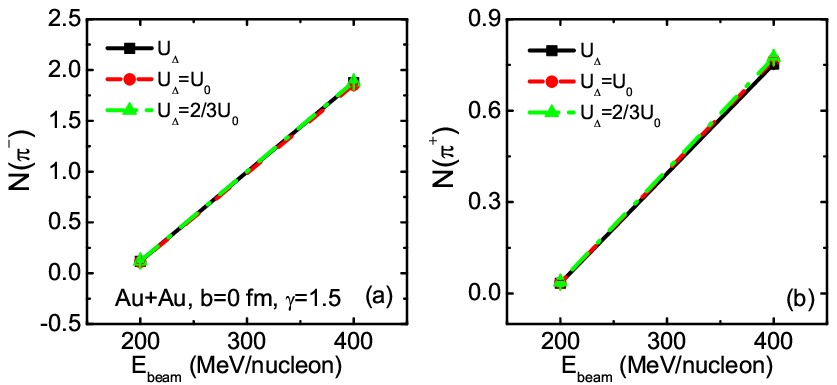}
\end{center}
\caption{(Color online) Same as Figure~\ref{g03-Npi}, but with stiff symmetry potential ($\gamma=1.5$).} \label{g15-Npi}
\end{figure}
\begin{figure}[htb]
\begin{center}
\includegraphics[width=0.5\textwidth]{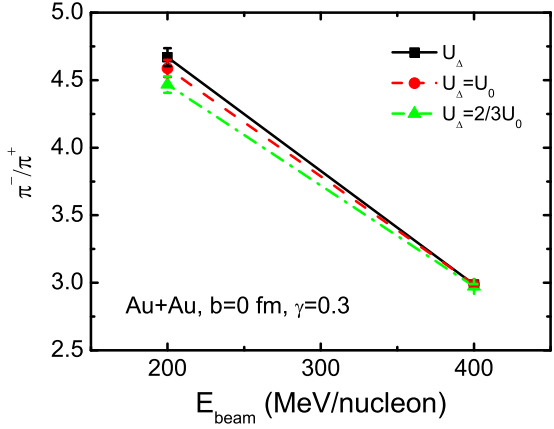}
\end{center}
\caption{(Color online) Effects of different forms of $\Delta$ resonance potential on $\pi^-/\pi^+$ ratio with soft symmetry potential ($\gamma=0.3$) in $^{197}Au+^{197}Au$ reaction at beam energies of 200 and 400MeV/A, respectively.} \label{gamma03}
\end{figure}
\begin{figure}[htb]
\begin{center}
\includegraphics[width=0.5\textwidth]{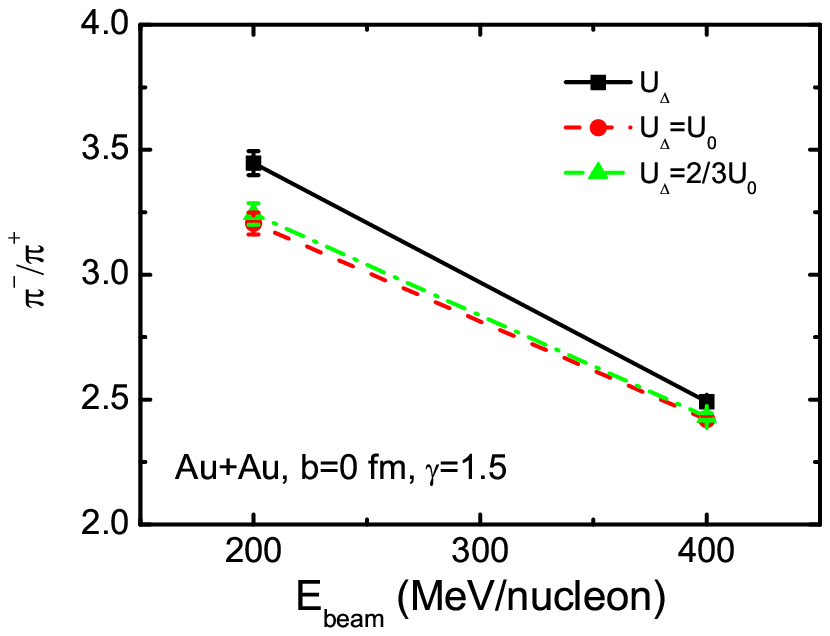}
\end{center}
\caption{(Color online) Same as Figure~\ref{gamma03}, but with stiff symmetry potential ($\gamma=1.5$).} \label{gamma15}
\end{figure}
\begin{figure}[htb]
\begin{center}
\includegraphics[width=0.5\textwidth]{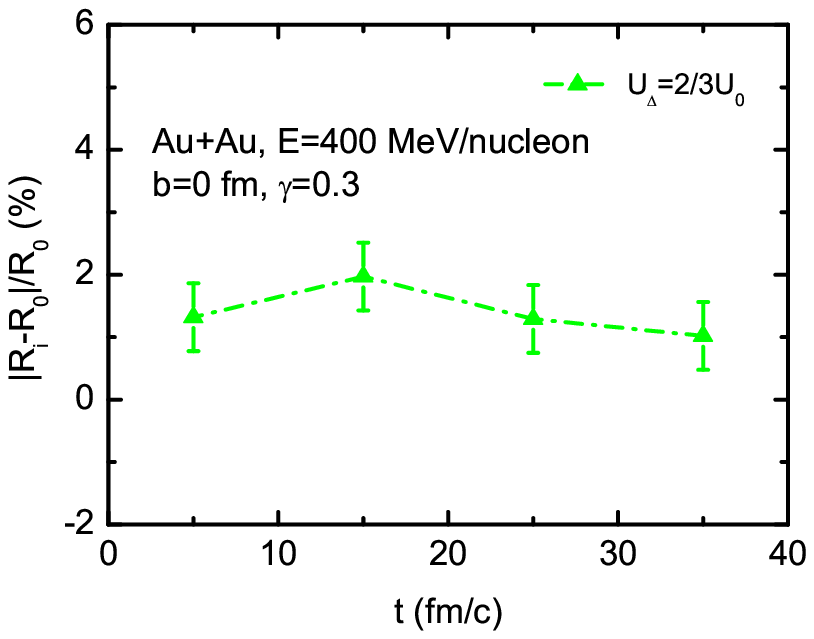}
\end{center}
\caption{(Color online) Relative sensitivity of the $\Delta$ resonance potential to the $\pi^-/\pi^+$ ratio in different reaction periods in $^{197}Au+^{197}Au$ reaction at a beam energy of 400MeV/A.} \label{piratio}
\end{figure}
Before studying the effect of $\Delta$ resonance potential on charged pion production, it is instructive to see if $\Delta$ resonance potential affects pre-equilibrium free $n/p$ ratio.
Fig.~\ref{npt} shows the effect of $\Delta$ resonance potential on pre-equilibrium free $n/p$ ratio. It is seen that whether for soft or stiff symmetry potentials, the form of $\Delta$ resonance potential does not affect pre-equilibrium free $n/p$ ratio at incident beam energies of 200 and 400 MeV/A. Only a small fraction of nucleons go through $\Delta$ resonances in the whole reaction process, one thus does not see any effect of $\Delta$ potential on pre-equilibrium free $n/p$ ratio \cite{LiBA15}.

Now we move to the study of pion production. Shown in Fig.~\ref{E400piEk} is kinetic energy dependence of $\pi^-/\pi^+$ ratio with different forms of $\Delta$ resonance potential and different symmetry potentials ($\gamma = 0.3,\  \gamma = 1.5$) and with or without pion potentials \cite{WMGuo15} in the central collision of $^{197}Au+^{197}Au$ at a beam energy of 400MeV/A. Fig.~\ref{E200piEk} is same as Fig.~\ref{E400piEk}, but for $E_{beam}$ = 200 MeV/A. 
Unlike the pion potential used in Ref.~\cite{hongj2014}, our pion potential includes the isoscalar part and the isovector part and both of them are momentum dependent \cite{WMGuo15}.
It is seen that the value of $\pi^-/\pi^+$ ratio with $\Delta$ resonance potential including isoscalar and iso-vector potentials is almost equal to that with only isoscalar $\Delta$ resonance potential within error. Around the Coulomb peak \cite{GCYong06}, there are small differences with different $\Delta$ resonance potentials, especially for different isoscalar potentials. Because differences of $\Delta$ resonance potentials with different charge states are in fact very small (as shown in Fig.~\ref{Deltapot}), one sees negligible effect of $\Delta$ isovector potential on the charged $\pi^-/\pi^+$ ratio. The general relative large discrepancy of the result corresponding $\frac{2}{3}U_0$ potential is due to its larger difference from the other two as shown in Fig.~\ref{Deltapot}. The shallower potential well of $\frac{2}{3}U_0$ potential causes $\Delta$ resonance has shorter time to exist in nuclear matter, thus less affected by the Coulomb potential. Therefore, around the Coulomb peak \cite{GCYong06}, the value of the $\pi^-/\pi^+$ ratio is not so large.

It is also instructive to see if $\Delta$ potential affects total charged pion production.
Shown in Fig.~\ref{g03-Npi} and Fig.~\ref{g15-Npi} are total charged pion productions at different incident beam energies with different $\Delta$ potentials and nucleon symmetry potentials. Again, one sees negligible effect of $\Delta$ potential on the charged $\pi$ production.
Fig.~\ref{gamma03} and Fig.~\ref{gamma15} show the effect of $\Delta$ resonance potential on $\pi^-/\pi^+$ ratio with soft and stiff symmetry potentials ($\gamma=0.3$, $\gamma=1.5$) at beam energies of 200 and 400MeV/A, respectively. One sees negligible effect of $\Delta$ potential on the charged $\pi$ production at higher incident beam energy $E_{beam}$ = 400 MeV/A whereas one may see observable effect (about 5\%) of $\Delta$ potential on the $\pi^-/\pi^+$ ratio at lower incident beam energy $E_{beam}$ = 200 MeV/A, especially for the stiff symmetry potential. The stiff symmetry potential causes larger difference among different charged (i.e., different isospin-dependent) $\Delta$ potentials. It is not surprise to see larger value of $\pi^-/\pi^+$ ratio with the isovector $\Delta$ potential.

Since $\Delta$ resonance lies in different reaction periods, it is thus
interesting to see in which reaction period the
$\Delta$ potential shows sensitivity to the $\pi^-/\pi^+$ ratio. In
order to know in which reaction period the $\pi^-/\pi^+$ ratio shows sensitivity to the
$\Delta$ potential, similar with previous studies \cite{liuy1,liuy2}, in
the whole reaction process $0 < t < 40$ fm/c we use
$U_{\Delta} = 0$ as the standard calculation, which gives a value of $R_{0}$, i.e.,
\begin{equation}
U_{\Delta}^{0 < t < 40 fm/c} \rightarrow R_{0}.
\end{equation}
To see the relative sensitivity in different
reaction periods (i.e., $t_{1} \leq 10 fm/c,~
10 fm/c < t_{2} \leq 20 fm/c, ~20 fm/c < t_{3} \leq 30%
fm/c, ~30 fm/c < t_{4} \leq 40 fm/c$), we turn on the $\Delta$ potential (e.g, $U_{\Delta} = 2/3U_{0}$) in one reaction period
but keep it null in the residual reaction period. We
thus get the other value of the $\pi^-/\pi^+$ ratio $R_{i}$, i.e.,
\begin{equation}
U_{\Delta}^{t_{i}} \rightarrow
R_{i} (i=1,2,3,4).
\end{equation}
By comparing these new computational results with the
standard calculation $R_{0}$, one can obtain the relative
sensitivity in a certain reaction period, i.e., $\frac{|R_{i}-R_{0}|}{R_{0}} \times 100$.
Shown in Fig.~\ref{piratio} is the decomposition of the sensitivity of the $\Delta$ resonance potential to the $\pi^-/\pi^+$ ratio in different reaction periods. It is seen that the effect of $\Delta$ resonance potential on the $\pi^-/\pi^+$ ratio in different reaction periods are not more than 2\%, thus the total effect of $\Delta$ resonance potential on the $\pi^-/\pi^+$ ratio
is roughly 5\%. As expected, the maximal effect of $\Delta$ resonance potential on the $\pi^-/\pi^+$ ratio is in the maximal compression period of the $^{197}Au+^{197}Au$ reaction.

\section{Conclusions}

In summary, in the framework of the isospin-dependent Boltzmann-Uehling-Uhlenbeck(IBUU) transport model, we studied the effects of $\Delta$ resonance potential on the free n/p and $\pi^-/\pi^+$ ratios in
$^{197}Au+^{197}Au$ reaction at beam energies of 200 and 400 MeV/A.
It is shown that the effect of $\Delta$ potential on the ratio of pre-equilibrium free $n/p$ is negligible. The effect of $\Delta$ potential on the ratio of $\pi^-/\pi^+$ may be observable only at lower incident beam energies. Relating to the question of non-conservation of energy on
$\Delta$ or $\pi$ productions in heavy-ion collisions, one can replace $\Delta$ potential by  nucleon isoscalar potential especially a soft symmetry energy is employed.

\section*{Acknowledgments}

This work is supported by
the National Natural Science Foundation of China (Grant Nos. 11375239, 11375062, 11175219), the 973 Program of China (Grant
No.2013CB834405), the Knowledge Innovation Project (Grant No. KJCX2-EW-N01) of Chinese Academy of Sciences, the project sponsored by
SRF for ROCS, SEM and the National Key Basic Research Program of China (No. 2013CB834400).

\end{document}